\documentstyle[12pt,titlepage]{article}
\begin{document}
\pagestyle{empty}
\baselineskip=0.212in

\begin{flushleft}
\large
{SAGA-HE-63-94
\hfill August, 1994}  \\
\end{flushleft}

\vspace{1.3cm}

\begin{center}

\Large{{\bf FORTRAN program for a numerical solution}} \\
\vspace{0.3cm}
\Large{{\bf of the nonsinglet Altarelli-Parisi equation}} \\

\vspace{1.0cm}

\Large
{R. Kobayashi, M. Konuma, and S. Kumano $^\star$ }         \\

\vspace{0.8cm}

\Large
{Department of Physics} \\

\vspace{0.1cm}

\Large{Saga University}    \\

\vspace{0.1cm}

\Large
{Saga 840, Japan}         \\

\vspace{1.0cm}

\Large{ABSTRACT}

\end{center}

We investigate a numerical solution of the flavor-nonsinglet
Altarelli-Parisi equation with next-to-leading-order $\alpha_s$
corrections by using Laguerre polynomials.
Expanding a structure function (or a quark distribution)
and a splitting function
by the Laguerre polynomials, we reduce
an integrodifferential equation to a summation of finite number
of Laguerre coefficients. We provide a FORTRAN program for
Q$^2$ evolution of nonsinglet structure functions
(F$_1$, F$_2$, and F$_3$) and nonsinglet quark distributions.
This is a very effective program with typical running
time of a few seconds on SUN-IPX or on VAX-4000/500.
Accurate evolution results are obtained by taking approximately
twenty Laguerre polynomials.

\vspace{1.5cm}

\noindent
{\rule{6.cm}{0.1mm}} \\

\vspace{-0.4cm}

\noindent
\normalsize
{$\star$ ~Email: kobar, konumam, or kumanos@himiko.cc.saga-u.ac.jp} \\

\vspace{0.0cm}
\hfill
{submitted for publication}

\vfill\eject
\pagestyle{plain}

\noindent
{\Large\bf {Program Summary}}
\vspace{0.4cm}

\medskip
\noindent
{\it Title of program:} LAG2NS

\medskip
\noindent
{\it Computer:} SUN-IPX (VAX-4000/500);
               {\it Installation:} Saga University Computer Center
                             (The Research Center for Nuclear Physics in Osaka)

\medskip
\noindent
{\it Operating system:} SUN-OS-4.1.2 (VAX/VMS V5.5-2)

\medskip
\noindent
{\it Programming language used:} FORTRAN 77

\medskip
\noindent
{\it Peripherals used:} laser printer

\medskip
\noindent
{\it No. of lines in distributed program, including test data, etc.:} 1111


\medskip
\noindent
{\it Nature of physical problem}

\noindent
This program solves the Altarelli-Parisi equation
for a spin-independent flavor-nonsinglet structure
function or quark distribution.

\medskip
\noindent
{\it Method of solution}

\noindent
We expand an initial quark distribution (or a structure function)
and a splitting function by Laguerre polynomials.
Then, the solution of the Altarelli-Parisi equation is expressed
in terms of the Laguerre expansion coefficients and the Laguerre polynomials.

\medskip
\noindent
{\it Restrictions of the program}

\noindent
This program is used for calculating Q$^2$ evolution of a spin-independent
flavor-nonsinglet structure function or quark distribution
in the leading order or in the next-to-leading-order of $\alpha_s$.
The double precision arithmetic is used.
The renormalization scheme is the modified minimal subtraction
scheme ($\overline{MS}$).
A user provides the initial structure function or
the quark distribution as a subroutine.
Examples are explained in sections 3.2, 4.13, and 4.14.
Then, the user inputs thirteen parameters explained in section 3.1.

\medskip
\noindent
{\it Typical running time}

\noindent
Approximately five (three) seconds on SUN-IPX (VAX-4000/500) if the initial
distribution is provided in the form of
$a_1x^{b_1}(1-x)^{c_1}+a_2x^{b_2}(1-x)^{c_2}+\cdot\cdot\cdot$ in the
subroutine GETFQN. If Laguerre coefficients of the initial
distribution in the subroutine FQNS(x)
are calculated by a GAUSS quadrature,
the running time becomes longer depending on the function form.

\vfill\eject
\vspace{1.5cm}
\noindent
{\Large\bf {LONG WRITE-UP}}

\vspace{1.0cm}
\noindent
{\Large\bf {1. Introduction}}
\vspace{0.4cm}

Structure functions measured in deep-inelastic
lepton-nucleon scattering depend in general on two kinematical
variables, $Q^2=-q^2$ and $x=Q^2/2P\cdot q$,
where $q$ is the four-momentum transfer and
P is the nucleon momentum.
If the nucleon is pointlike,
the structure functions depend only on the variable $x$.
This assumption is called Bjorken scaling hypothesis.
However, observed experimental data indicate weak Q$^2$ dependence,
and this fact is referred to as scaling violation.
Although the structure functions themselves
could not be calculated exactly in QCD except for lattice QCD methods,
the scaling violation can be evaluated
in perturbative QCD.
The scaling violations of the
structure functions have been investigated extensively, and
an intuitive way to describe the phenomena
is to use the Altarelli-Parisi equation \cite{AP}:
$$
{d\over d\ln Q^2} ~q_{_{NS}}(x,Q^2)
={\alpha_s(Q^2)\over 2\pi}\int_x^1{dy\over y}~
P_{_{NS}}\biggl({x\over y}\biggr)~
q_{_{NS}}(y,Q^2) ~~~,
\eqno{(1.1)}
$$
where $q_{_{NS}}(x,Q^2)$ is a flavor-nonsinglet quark distribution,
$\alpha_s(Q^2)$ is the running coupling constant in Appendix A,
and $P_{NS}(x)$ is the nonsinglet splitting function
in Appendix B.
The above equation describes the process that
a quark with the nucleon's momentum fraction $y$ radiates
a gluon and it becomes a quark with the momentum fraction $x$.
The splitting function $P_{NS}(z)$ determines
the probability for a quark
radiating a gluon such that the quark momentum is reduced
by the fraction $z$.
Next-to-leading-order (NLO) corrections to
the Altarelli-Parisi equation are included
in the coupling constant $\alpha_s$
and in the splitting function $P_{NS}$
$$
{d\over d\ln Q^2} ~q_{_{NS}}(x,Q^2)={\alpha_s(Q^2)\over 2\pi}
\int_x^1{dy\over y} ~\biggl[P_{_{NS}}^{(0)}
\biggl({x\over y}\biggr)+{\alpha_s(Q^2)\over 2\pi}
  P_{_{NS}}^{(1)}\biggl({x\over y}\biggr)\biggr] ~
q_{_{NS}}(y,Q^2)  ~,
\eqno{(1.2)}
$$
where the running coupling constant is
the NLO expression $\alpha_s^{NLO}$ in Eq. (A.2),
and the leading order (LO) and NLO splitting functions
are given in Eqs. (B.1), (B.3), and (B.5).
The leading order and the next-to-leading order are abbreviated to
LO and NLO respectively throughout this paper.

Because the above Q$^2$ evolution equation is very important for testing
the perturbative QCD and is often used theoretically and experimentally,
it is worth while having a computer program of solving it
accurately without consuming much computing time.
There are several methods for solving the above integrodifferential
equation. These include brute-force methods \cite{BRUTE},
Mellin-transformation methods \cite{MELLIN},
methods of using orthogonal polynomials \cite{MELLIN},
and others.
Among these, the Laguerre method in Refs. \cite{FP,RAM,KL}
is considered to be very effective.
In Ref. \cite{KL}, we provide the Laguerre method program LAG1,
which deals with the Q$^2$ evolution in the leading order.
The purpose of this paper is to extend the nonsinglet program
in Ref. \cite{KL} by including the NLO corrections.

In section 2, we explain the Laguerre method for solving the evolution
equation with the NLO corrections. In section 3, we supply necessary
information for running our program LAG2NS.
The details of each subroutine are discussed in section 4, and
numerical results are explained in section 5.
Explicit expressions for the running coupling constant, the splitting
function, the Laguerre polynomials, and other necessary functions
are given in appendices.

\vspace{1.0cm}
\noindent
{\Large\bf {2. Laguerre polynomial method}}
\vspace{0.4cm}

The Laguerre polynomial method
was first investigated by Furmanski and Petronzio \cite{FP}.
The quark distribution and the splitting function are
expanded in the Laguerre polynomials, then the solution
is given by the Laguerre expansion coefficients and
the Laguerre polynomials.
Numerical analysis of the method was studied by Ramsey \cite{RAM}.
Especially, the fast-computing Laguerre method was developed
if the initial distribution is expressed
in a simple analytical form $a x^b (1-x)^c$.
According to Ref. \cite{RAM}, the Laguerre polynomial method is
considered to be more effective than other methods in convergence and
in computing time.
This is because the first several polynomials resemble
parton distributions.
Based on these investigations, a useful FORTRAN program
was published in Ref. \cite{KL}.
However, the program can be used only for the evolution
in the leading order of $\alpha_s$.
We discuss the solution for the evolution equation including
the NLO corrections in the following. We refer the reader to the papers
in Refs. \cite{FP,RAM,KL} for more complete account of
the Laguerre method.

We outline the Laguerre method with the NLO corrections.
In the following, we use the NLO running coupling constant
unless we specify LO. We use $\overline{MS}$ as the renormalization
scheme.
Because the quark distribution multiplied by $x$
satisfies the same evolution equation,
we define $\widetilde q(x)=xq(x)$ and $\widetilde P(x)=xP(x)$
and rewrite Eq. (1.2) as
$$
{d\over d\ln Q^2} ~ \widetilde q_{_{NS}}(x,Q^2)={\alpha_s(Q^2)\over 2\pi}
\int_x^1{dy\over y} ~
\biggl[\widetilde P_{_{NS}}^{(0)}\biggl({x\over y}\biggr)
+{\alpha_s(Q^2)\over 2\pi}
  \widetilde P_{_{NS}}^{(1)}\biggl({x\over y}\biggr)\biggr] ~
\widetilde q_{_{NS}}(y,Q^2)  ~.
\eqno{(2.1)}
$$
In order to remove the Q$^2$ dependence in front of the integral,
we use the variable $t$ defined by
$$
t \equiv -{2\over \beta_0}\ln
\biggl[{\alpha_s(Q^2)\over \alpha_s(Q_0^2)
}\biggr]    ~~~,
\eqno{(2.2)}
$$
instead of $Q^2$, where the running coupling constant $\alpha_s$
is given in Eq. (A.2) and the constant $\beta_0$ is in Eq. (A.3).
For example, $Q^2$ evolutions from $Q_0^2$=4 GeV$^2$ to
$Q^2$=10, 100, and 1000 GeV$^2$ with $\Lambda=0.2$ GeV
and four flavors correspond to
$t=$0.00384, 0.112, and 0.166.
Then, the integrodifferential equation becomes simpler
$$
{d\over dt} ~ \widetilde q_{NS}(x,t)=
\int_x^1{dy\over y} ~
\biggl[\widetilde P_{NS}^{(0)}\biggl({x\over y}\biggr)
+{\alpha_s(t)\over 2\pi}
  R\biggl({x\over y}\biggr)\biggr]  ~
\widetilde q_{NS}(y,t) ~~~,
\eqno{(2.3)}
$$
where $R(x)$ is given by $\widetilde P^{(0)}(x)$
and $\widetilde P^{(1)}(x)$ as
$$
R(x)\equiv \widetilde{P}_{NS}^{(1)}(x)-{\beta_1\over 2\beta_0}
\widetilde{P}_{NS}^{(0)}(x)  ~~~.
\eqno{(2.4)}
$$
The constant $\beta_1$ is given in Eq. (A.3).

We define the evolution operator $E(x,t)$ by the convolution
form
$$
\widetilde{q}_{_{NS}}(x,t) =
\int_x^1{dy\over y} ~E({x\over y},t) ~\widetilde{q}_{_{NS}}(y,t=0) ~~~.
\eqno{(2.5)}
$$
Namely, the function $E(x,t)$ describes the Q$^2$ evolution
from the initial distribution at Q$_0^2$ ($t=0$)
to the distribution at Q$^2$. From Eqs. (2.3) and (2.5),
it satisfies the integrodifferential equation
$$
{d\over dt} ~ E(x,t)=\int_x^1{dy\over y}
+{\alpha_s(t)\over 2\pi}R({x\over y})\biggr]~ E(y,t)  ~~~.
\eqno{(2.6)}
$$
Because Eq. (2.6) and Eq. (2.3) are exactly the same form, it may seem
unnecessary to define such operator. The introduction of the evolution
operator simplifies the following formalism because it is the $\delta$
function at $t=0$, $E(x,t=0)=\delta (1-x)$.
Laguerre coefficients of $E(x,t=0)$ are calculated by using Eq. (C.3),
and they are $E_n(t=0)=1$ for all $n$.
We write the function $E(x,t)$ by LO and NLO contributions as
$$
E(x,t)=E^{(0)}(x,t)+{\alpha_s(0)\over 2\pi}E^{(1)}(x,t) ~~~.
\eqno{(2.7)}
$$
We substitute this equation into Eq. (2.6). Then,
considering that the LO evolution operator $E^{(0)}(x,t)$
satisfies the LO Altarelli-Parisi equation
$$
{d\over dt} ~ E^{(0)}(x,t)=
\int_x^1{dy\over y} ~\widetilde{P}_{NS}^{(0)}({x\over y}) ~E^{(0)}(y,t)  ~~~,
\eqno{(2.8)}
$$
we find that the NLO operator satisfies the equation
$$
{d\over dt} ~E^{(1)}(x,t)={{\alpha_s(t)} \over {\alpha_s(0)}}
\int_x^1{dy\over y}~ R({x\over y}) ~E^{(0)}(y,t)
+\int_x^1{dy\over y} ~\widetilde{P}_{NS}^{(0)}({x\over y}) ~E^{(1)}(y,t)  ~~~.
\eqno{(2.9)}
$$
This equation indicates that the LO and NLO evolution operators
are not independent and they
are related by \cite{MISTAKE}
$$
 E^{(1)}(x,t)={2\over \beta_0}\biggl[1-{\alpha_s(t) \over \alpha_s(0)}\biggr]
\int_x^1 {{dy} \over y}  ~R\biggl({x \over y}\biggr)~ E^{(0)}(y,t)  ~~~.
\eqno{(2.10)}
$$
So we do not have to solve equations for
both $E^{(0)}(x,t)$ and $E^{(1)}(x,t)$.
We solve Eq. (2.8) for getting $E^{(0)}(x,t)$,
which is then substituted into
Eq. (2.10) for calculating $E^{(1)}(x,t)$.

We introduce variables $x'$ and $y'$ by
$x^\prime \equiv -\ln x$ and $y^\prime \equiv -\ln y$,
where $0\le x, y\le 1$,
because the Laguerre polynomials are defined in the region
$0\le x'< \infty$.
Using these variables,
we expand the functions
$E(x=e^{-x^\prime})$ and $\widetilde{P}(x=e^{-x^\prime})$
in terms of the Laguerre polynomials,
$\displaystyle{f (e^{-x^\prime})=\sum_n f_n L_n(x^\prime)}$.
Then our problem is to obtain the expansion coefficients
$E_n$ in terms of the coefficients of the splitting function $P_n$.
The details of solving Eq. (2.8) are discussed
in Ref. \cite{KL} as well as in Refs. \cite{FP,RAM},
and we summarize the solution in Appendix C.
The NLO evolution operator is then given by
using the calculated $E_n^{(0)}(t)$
$$
E_n^{(1)}={2\over \beta_0}
\biggl[ 1- {{\alpha_s(t)} \over {\alpha_s(0)}} \biggr]
\widehat E_n^{(1)}(t)   ~~~,
\eqno{(2.11)}
$$
where $\widehat E_n^{(1)}(t)$ is defined by
$$
\widehat E_n^{(1)}(t)=\sum_{i=0}^n(R_{n-i}-R_{n-i-1})E_i^{(0)}(t) ~~~.
\eqno{(2.12)}
$$
$R_i$ is the Laguerre coefficients of $R(x)$.
$R_{-1}$ is defined as $R_{-1}=0$.
Using the evolution operators in Eqs. (C.5) and (2.11),
we finally obtain
$$
\widetilde{q}_{_{NS}}(x,t)=\sum_{n=0}^N \sum_{m=0}^n
\biggl[E_{n-m}(t)-E_{n-m-1}(t)\biggr]
L_n(- \ln x)\widetilde{q}_{_{NS},m}(t=0) ~~~,
\eqno{(2.13)}
$$
with $\displaystyle{E_n(t)=E_n^{(0)}(t)
      +{{\alpha_s(0)} \over {2\pi}} E_n^{(1)}(t)}$.
In this way, we reduced the original integrodifferential equation
to a sum of finite number of Laguerre expansion coefficients
and the Laguerre polynomials.
If the Laguerre coefficients of the splitting functions
and the initial distribution
are calculated fast,
the Altarelli-Parisi equation is solved numerically
without consuming much computing time.

We have discussed a solution for the $Q^2$ evolution of
a nonsinglet quark distribution.
$Q^2$ evolution of a nonsinglet structure function
is calculated in a similar way.
The only modification is to
take into account NLO corrections
from the coefficient function.
A nonsinglet structure function $F_{NS}(x,Q^2)$ is expressed
as a convolution of
the corresponding nonsinglet quark distribution $q_{_{NS}} (x,Q^2)$
and the coefficient function $C_{NS} (x,\alpha_s)$ as \cite{HW,RGR}
$$
F_{NS}(x,Q^2)=\int_x^1 {{dy} \over y} ~ C_{NS} ({x \over y},\alpha_s)
                                      ~ q_{_{NS}} (y,Q^2)  ~~~.
\eqno{(2.14)}
$$
The coefficient function is
$$
C_{NS} (x,\alpha_s)=\delta (1-x)
                    +{{\alpha_s} \over {4\pi}} B^{NS}(x)   ~~~.
\eqno{(2.15)}
$$
$B^{NS}(x)$ is the NLO correction, and those
for the structure functions
($F_1$, $F_2/x$, and $F_3$) are listed in Eq. (B.6).
Combining the evolution equation for the quark distribution
in Eq. (1.2) with the convolution
in Eq. (2.14), we obtain the $Q^2$ evolution equation for
the nonsinglet structure function as \cite{HW}
$$
{d \over d\ln Q^2} F_{NS}(x,Q^2)={{\alpha_s(Q^2) }\over 2\pi}
\int_x^1 {dy \over y}
\biggl[P_{NS}^{(0)} ({x \over y})
+{{\alpha_s(Q^2)}\over 2\pi}
\biggl(P_{NS}^{(1)} ({x \over y})
-{1 \over 4}\beta_0 B^{NS}({x \over y}) \biggr)\biggr] F_{NS}(y,Q^2)   ~.
\eqno{(2.16)}
$$
This is exactly the same equation with Eq. (1.2) if we replace
$P_{NS}^{(1)} (x)$ by
$P_{NS}^{(1)} (x)-{1 \over 4}\beta_0 B^{NS} (x)$.
Therefore, we do not have to solve the above
equation for $F_{NS}(x,Q^2)$ independently.
The solution Eq. (2.13) can be used with the replacements
$$
\widetilde q_{_{NS}}(x,Q^2)\rightarrow
            x F_1(x,Q^2), ~~ F_2(x,Q^2), ~~ or~~ x F_3(x,Q^2) ~~~,
\eqno{(2.17a)}
$$
and
$$
R(x)
\rightarrow
\widetilde{P}_{NS}^{(1)}(x)
-{1 \over 4}\beta_0 B_n^{NS}(x)
-{\beta_1 \over{2\beta_0}}\widetilde{P}_{NS}^{(0)}(x), ~~~n=1,~2,~ or~3 ~~~.
\eqno{(2.17b)}
$$

In the following sections, we explain our program of calculating
Eq. (2.13) or the corresponding equation for the structure function.

\vfill\eject
\begin{tabbing}
{\Large\bf 3} $~$ \= {\Large\bf Description of input parameters}       \\
$~~~$             \> {\Large\bf and input distribution}
\end{tabbing}
\vspace{0.4cm}

\noindent
{\it 3.1 Input parameters}
\vspace{0.4cm}

There are thirteen input parameters in the main program.

$~~~$

\noindent
\begin{tabular}{rllll}
 (1) & F0      & \multicolumn{3}{l}
                {= number of quark flavors (F0=3 or 4).}                    \\
 (2) & IORDER  & = 1, & \multicolumn{2}{l}
                        {leading order in $\alpha_ s$;}                      \\
     &         & = 2, & \multicolumn{2}{l}
                        {next-to-leading order.}                             \\
 (3) & ITYPE   & = 1, & structure function &$xF_1^{^{NS}}(x,Q^2)$;       \\
     &         & = 2, &                    &$~~F_2^{^{NS}}(x,Q^2)$;       \\
     &         & = 3, &                    &$xF_3^{^{NS}}(x,Q^2)$;      \\
     &         & = 4, & \multicolumn{2}{l}
                        {quark distribution $xq_{_{NS}}(x,Q^2)$.}          \\
 (4) & IMORP   & = 1, & \multicolumn{2}{l}
                        {$q-\bar q$ type distribution;}
\\
     &         & = 2, & \multicolumn{2}{l}
                        {$q+\bar q$ type distribution.}
\\
 (5) & Q02     & \multicolumn{3}{l}
                {= initial $Q^2$ ($\equiv Q_0^2$)
                   at which an initial distribution is supplied.}           \\
 (6) & Q2      & \multicolumn{3}{l}
                {= $Q^2$ (in GeV$^2$) to which the distribution is evolved.}
\\
 (7) & DLAM    & \multicolumn{3}{l}
                {= QCD scale parameter $\Lambda_{QCD}$ in GeV.}            \\
 (8) & NMAX    & \multicolumn{3}{l}
                {= maximum order
                    of Laguerre polynomials\ (0$ < $NMAX$ < $31).}         \\
 (9) & IQN     & = 1, & \multicolumn{2}{l}
                        {calculate the Laguerre coefficients of
                        the initial distribution by}                         \\
     &         &      & \multicolumn{2}{l}
                        {Eq. (C.4);}                                        \\
     &         & = 2, & \multicolumn{2}{l}
                        {calculate them by a Gauss-Legendre quadrature.}
\\
(10) & ILOG    & = 1, & \multicolumn{2}{l}
                        {linear scale in $x$;}
\\
     &         & = 2, & \multicolumn{2}{l}
                        {logarithmic scale in $x$.}
\\
(11) & XMIN    & \multicolumn{3}{l}
                {= minimum of $x$ ($x \ne 0$).}
        \\
(12) & XMAX    & \multicolumn{3}{l}
                {= maximum of $x$.}                                         \\
(13) & NXSTEP  & \multicolumn{3}{l}
                {= number of $x$ at which
                     distributions are calculated.}
\end{tabular}

$~~~$

The above ILOG, XMIN, XMAX, and NXSTEP are parameters for writing
evolution results as the function of $x$. For example, the evolved
distributions are calculated at x=0.1, 0.2, ..., 1.0 for
ILOG=1, XMIN=0.1, XMAX=1.0, and NXSTEP=9.
XMIN=0 should be avoided due to a numerical problem.
In the end of this paper, a test run result is given for the parameter
set: F0=4.0, IORDER=2, ITYPE=4, IMORP=1, Q02=4.0, Q2=200.0, DLAM=0.19,
NMAX=30, IQN=1, ILOG=2, XMIN=0.01, XMAX=1.0, and NXSTEP=30.

\vspace{0.6cm}
\noindent
{\it 3.2 Input distribution}
\vspace{0.4cm}

If the initial distribution is written as
$a_1 x^{b_1} (1-x)^{c_1} + a_2 x^{b_2} (1-x)^{c_2} +\cdot\cdot\cdot$,
the input IQN=1 should be chosen in order to save computing time.
In this case, the subroutine GETFQN should be supplied
to calculate Laguerre coeficients of the
input distribution at $Q_0^2$, $ [xq_{_{NS}}(x)] _n $
or $[xF_{NS}(x)]_n$.
In this subroutine, the {\it n}th Laguerre coefficient
for the initial distribution
$ ax^b(1-x)^c $ is calculated by calling the function QN(n,a,b,c).
An example is given in Sec. 4.14.

If the initial distribution cannot be written in the simple form,
the input IQN=2 should be chosen.
Then, the initial distribution at $Q_0^2$
should be supplied as the double precision function
FQNS($x$).
An example is given in Sec. 4.13.

\vspace{1.0cm}
\noindent
{\Large\bf {4. Description of the program LAG2NS}}
\vspace{0.4cm}

\noindent
{\it 4.1 Main program LAG2NS}
\vspace{0.3cm}

The main program reads thirteen input parameters
in section 3.1 from the input file 5.
$t$ defined in Eq. (2.2) is calculated in LO and in NLO.
Then, the subroutine GETFQNS is called for
calculating the $Q^2$ evolution.

\vspace{0.4cm}
\noindent
{\it 4.2 Functions ALPHA0(Q2) and ALPHA1(Q2)}
\vspace{0.3cm}

Running coupling constants in LO and NLO
are given in ALPHA0 and ALPHA1 respectively.
The NLO $\alpha_s$ is given in the $\overline {MS}$ scheme.
See Appendix A for the details.

\vspace{0.4cm}
\noindent
{\it 4.3 Subroutines GETFQNS, OBTAIN1, OBTAIN2, and OBTAINF}
\vspace{0.3cm}

The subroutine GETFQNS calculates $Q^2$ evolution of a nonsinglet distribution.
First, the subroutine OBTAIN1 or OBTAIN2
(OBTAINF in the structure function case) is called,
depending on the input parameter IORDER.
These subroutines call all special functions used in the program.
If IORDER=1, the subroutine OBTAIN1 calls necessary functions for
calculating the LO evolution.
Then, the evolved distribution in the leading order is calculated
by calling the function FQNS0$(x)$
for Bjorken $x$ from XMIN to XMAX.
The evolved distribution is written on the output file 6.
If IORDER=2, OBTAIN2 (or OBTAINF) is called and
the evolved distribution including the NLO corrections
is calculated in the same way
by calling the function FQNS1$(x,t)$.
The subroutine OBTAIN2 (OBTAINF) calls necessary functions for
calculating NLO evolution results.

\vspace{0.4cm}
\noindent
{\it 4.4 Subroutines GETFACT, GETZETA, and GETZFUN}
\vspace{0.3cm}

These are necessary functions for calculating the LO evolution.
See Ref. \cite{KL} for the details.

\vspace{0.4cm}
\noindent
{\it 4.5 Subroutine GETPN0}
\vspace{0.3cm}

This subroutine calculates Laguerre coefficients of
the splitting function in the leading order;
coefficients are stored in the array PNSNN$ (n+1) [n=0-NMAX] =(xP_{NS})_n $.
Then $ P_n-P_{n-1} $ is calculated and
results are stored in the array PTILNSN$\  (n+1)\ [n=0-NMAX]$ .
These are used for calculating $ B_n^k $ in the subroutine GETBKM1.

\vspace{0.4cm}
\noindent
{\it 4.6 Subroutines GETRN3M, GETRN4M, GETRN3P, and GETRN4P.}
\vspace{0.3cm}

These subroutines store Laguerre coefficients
of the splitting function $R(x)$ [Eq. (2.4)]
in the array RN$ (n+1)\ [n=0-NMAX]$.
These coefficients are numerically calculated in
a separate program.
GETRN3M is for evolution of a ``$q-\bar q$" type distribution
in the three-flavor case, and
GETRN4M is in the four-flavor case.
GETRN3P and GETRN4P are for a ``$q+\bar q$" type distribution
with three and four flavors respectively.
These are used for calculating $ \hat E_n^{(1)} $
[Eq. (2.12)] in the subroutine GETEMT1.
See Ref. \cite{HW} for the explanation of the
``$q-\bar q$" and ``$q+\bar q$" type distributions.

\vspace{0.4cm}
\noindent
{\it 4.7 Subroutines GETRN and GETFN }
\vspace{0.3cm}

In these subroutines, $ R_n-R_{n-1} $ is calculated and
results are stored in the array RTILN$\  (n+1)\ [n=0-NMAX]$ .
These are used for calculating $\widehat E_n^{(1)}$ [Eq. (2.12)]
in the subroutine GETEMT1.
In the case of structure-function evolution,
$R(x)$ includes the coefficient function $B^{NS}(x)$
as shown in Eq. (2.17b).
$R(x)$ defined in Eq. (2.4) is used in GETRN, and
Eq. (2.17b) is used in GETFN for calculating $R_n-R_{n-1}$.

\vspace{0.4cm}
\noindent
{\it 4.8 Subroutines GETF1N, GETF2N, and GETF3N}
\vspace{0.3cm}

Laguerre coefficients of the function $B^{NS}(x)$
are supplied in these subroutines.
These coefficients are numerically calculated in
a separate program. The functions $B^{NS}(x)$
for the structure functions, $F_1$, $F_2/x$, and $F_3$,
are given in Eq. (B.6) and
their (actually $B^{NS}(x)$ multiplied by $x$) Laguerre coefficients
are supplied in GETF1N, GETF2N, and GETF3N respectively.
The Laguerre coefficients are stored in the array
FNN$\  (n+1)\ [n=0-NMAX]$.

\vspace{0.4cm}
\noindent
{\it 4.9 Subroutine GETBKM1}
\vspace{0.3cm}

This subroutine calculates $B_m^k$ in Eqs. (C.6) and (C.7),
then they are used for calculating the LO evolution
operator in Eq. (C.5).

\vspace{0.4cm}
\noindent
{\it 4.10 Function QN(N,a,b,c) and subroutine GETQND}
\vspace{0.3cm}

QN(N,a,b,c) is used for calculating Laguerre coefficients
of the initial distribution $a x^b (1-x)^c$
in the subroutine GETFQN in the end of this program.

GETQND calculates Laguerre coefficients by using
the Gauss quadrature subroutine DGAUSS with the accuracy
of $10^{-8}$.

\vfill\eject
\noindent
{\it 4.11 Subroutines GETEMT0 and GETEMT1}
\vspace{0.3cm}

GETEMT0 calculates Laguerre coefficients
of the evolution operator $ E_n^{(0)}(t) $ in Eq. (C.5).
In calculating this equation, we need $ [xP_{NS}]_0 $
calculated in the subroutine GETPN0 and $ B_m^k $
in the subroutine GETBKM1.

To calculate the evolution of a quark distribution,
the differences $ E_n^{(0)}(t)-E_{n-1}^{(0)}(t) $
are calculated; the resulting differences are stored in the array
EPSNS$ (n+1)[n=0-NMAX] $.

GETEMT1 calculates Laguerre coefficients
of the evolution operator $ \hat E_n^{(1)}(t) $ in Eq. (2.12).
To calculate this, RTILN calculated in subroutine GETRN
(GETFN) and EPSNS are required.
Then, the differences $ \hat E_n^{(1)}(t) - \hat E_{n-1}^{(1)}(t) $
are calculated and the results are stored
in the array EPSNS1$ (n+1)[n=0-NMAX] $.
They are used for calculating the evolution of a quark distribution
in the next-to-leading order.

\vspace{0.4cm}
\noindent
{\it 4.12 Functions FQNS0(x) and FQNS1(x)}
\vspace{0.3cm}

The evolved distribution in the LO (in the NLO)
is calculated by using Eq. (2.13)
in the function FQNS0 (FQNS1).
It is the quark distribution
$FQNS0=x q_{_{NS}}(x,Q^2)$ for the input parameter ITYPE=4, and
it is the structure function $FQNS0=F_{NS}(x,Q^2)$
[$xF_1^{NS}$, $F_2^{NS}$, or $xF_3^{NS}$]
for ITYPE=1, 2, or 3.

\vspace{0.4cm}
\noindent
{\it 4.13 Function FQNS(x)}
\vspace{0.3cm}

If the initial distribution cannot be expressed in the simple form
$a x^b(1-x)^c$
(IQN=2), the distribution should be supplied as the function FQNS(x).
The HMRS-B valence-quark distribution $xu_v(x)+xd_v(x)$ \cite{HMRS} and
the CCFR $F_3$ structure function $xF_3(x)$ \cite{CCFR}
are given as examples
$$
[xu_v+xd_v]^{HMRS-B} = 0.5469 x^{0.237} (1-x)^{4.07}(1+23.8x) ~~~,
\eqno{(4.1)}
$$
$$
[xF_3]^{CCFR} = 5.976 x^{0.766} (1-x)^{3.101} ~~~.
\eqno{(4.2)}
$$

\vspace{0.4cm}
\noindent
{\it 4.14 Subroutine GETFQN: This subroutine or the function
                            $FQNS(x)$ should be supplied by a user.}
\vspace{0.3cm}

If the initial distribution can be expressed in the form
$a_1 x^{b_1}(1-x)^{c_1}+a_2 x^{b_2}(1-x)^{c_2}+\cdot\cdot\cdot$
(IQN=1),
the Laguerre coefficients are calculated by using the function
QN(N,A,B,C) as $QN(N,a_1,b_1,c_1)+QN(N,a_2,b_2,c_2)+\cdot\cdot\cdot$.
The HMRS-B and CCFR distributions are given as examples
$$
[xu_v+xd_v]_n^{HMRS-B} = QN(n,0.5469,0.237,4.07)
                        +QN(n,13.016,1.237,4.07) ~~~,
\eqno{(4.3)}
$$
$$
[xF_3]_n^{CCFR} = QN(n,5.976,0.766,3.101) ~~~.
\eqno{(4.4)}
$$

\vfill\eject
\vspace{1.0cm}
\noindent
{\Large\bf {5. Numerical analysis}}
\vspace{0.4cm}

Our FORTRAN-77 program LAG2NS can be run in double precision
arithmetic. This is a program for the $Q^2$ evolution
of a spin-independent flavor-nonsinglet structure function
or quark distribution with or without the NLO corrections.

The initial distribution
or its Laguerre coefficients should be
provided by the function
FQNS($x$) or the subroutine GETFQN
within the program.
If the initial distribution can be expressed in the analytical form
$a_1x^{b_1}(1-x)^{c_1}+a_2x^{b_2}(1-x)^{c_2}+\cdot\cdot\cdot$,
a user selects the input IQN=1 and provides the subroutine GETFQN.
In this case, computing time is very short.
However, if the distribution cannot be expressed in
the simple analytical form,
the function subroutine FQNS($x$) should be supplied.
Our program calculates Laguerre coefficients of this
distribution by a GAUSS quadrature with the accuracy of $10^{-8}$.
So it could take a significant amount of computing time
depending on complexity of the initial function.
Then, thirteen input parameters should be provided
from the input file 5 for running
the program.
Evolution results are written in the output file 6.
An example of the output is shown in the end
of this paper as TEST RUN OUTPUT.

We tested our program in comparison with other publications.
In order to save computing time, the Laguerre coefficients of
the NLO splitting functions and the coefficient functions are
calculated in separate programs.
Obtained Laguerre coefficients are supplied in our program
as discussed in sections 4.6 and 4.8.
In the separate programs,
the splitting-function and coefficient-function
subroutines are tested by calculating their moments,
namely anomalous dimensions.
Calculated results are compared with those in Refs. \cite{FRS,BBDM}.
Because the NLO expressions in the appendix B are rather complicated,
the Laguerre coefficients are evaluated numerically by the Gauss quadrature
with the accuracy of 10$^{-12}$.
We compared our NLO evolution results of $u_v(x,Q^2)+d_v(x,Q^2)$
with those of the HMRS-B program \cite{HMRS}.
The HMRS-B distribution at $Q^2$=4 GeV$^2$
is evolved to $Q^2$=200 GeV$^2$ with four flavors and
$\Lambda_{\overline{MS}}$=0.19 GeV.
We find that differences between
our evolution and the HMRS one are of the order of
0.5\% in the $x$ range $0.01<x<0.5$.
The difference becomes slightly larger in the large $x$ region; however,
it is still of the order of 2\%.
Hence, our program results essentially agree with
the HMRS $Q^2$ evolution.
We also checked our NLO corrections in comparison with
Ref. \cite{HWW}. Namely, we choose the starting nonsinglet $F_2$
structure function $F_2(x,Q^2=5 GeV^2)=3.8075x^{0.56}(1-x)^{2.71}$.
Then the structure function at $Q^2=200$ GeV$^2$ is calculated
with $\Lambda$=0.3 GeV and four flavors.
The results agree with those in Fig. 1b of Ref. \cite{HWW}; however,
we find a slight deviation in the medium $x$ region.

Next, we show examples of our evolution results.
We choose the HMRS-B distribution at $Q^2$=4 GeV$^2$
[$u_v+d_v=0.5469x^{0.237}(1-x)^{4.07}(1+23.8x)$]
as the initial distribution.
This distribution is evolved to the one
at $Q^2$=200 GeV$^2$ by assigning the input parameters
F0=4.0, IORDER=2, ITYPE=4, IMORP=1, Q02=4.0, Q2=200.0,
DLAM=0.19, NMAX=10, 20, or 30, IQN=1, ILOG=2, XMIN=0.01, XMAX=1.0,
and NSTEP=50 and by providing the subroutine GETFQN.
Evolved valence-quark distributions are shown in Fig. 1.
We find that reasonably accurate results are obtained by
taking about twenty Laguerre polynomials.
The accuracy is within 3\% in the $x$ range $0.001<x<0.8$.
However, there is a tendency to become worse in the very small
$x$ region ($x<0.001$) and in the very large $x$ region ($x>0.9$).

We compare our evolution results with the CDHSW
neutrino data \cite{NEUTRINO} in Fig. 2.
We choose the CCFR $F_3(x)$ distribution at $Q^2$=3 GeV$^2$
as the initial distribution
in both LO and NLO cases in order to see NLO contributions.
It is given by $xF_3(x,Q^2=3~GeV^2)=5.976x^{0.766}(1-x)^{3.101}$.
We calculate $Q^2$ variations with $\Lambda$=0.21 GeV,
four flavors, and the $q-\bar q$ splitting function
by providing the subroutine GETFQN.
The $Q^2$ variations are calculated at $x$=0.045, 0.225, and 0.55.
The NLO contributions are important at small $Q^2$ ($\approx$ 1 GeV$^2$)
and at small $x$ as shown in Fig. 2.
Our $Q^2$ variations are consistent with the CDHSW experimental data.

Typical running time is a few seconds on SUN-IPX or on VAX-4000/500.
Considering the accuracy and the short running time, we
conclude that the Laguerre method is considered to be
very effective for numerical solution of the Altarelli-Parisi
equation.

We have investigated a numerical solution of spin-independent
flavor-nonsinglet Altarelli-Parisi equation.
We will work on flavor-singlet structure functions
and also on nuclear structure functions.

\vspace{1.0cm}
\noindent
{\Large\bf Acknowledgements}
\vspace{0.4cm}

We would like to thank Drs. P. J. Mulders and A. W. Schreiber
for their suggestions.
We thank the Research Center for Nuclear Physics
in Osaka for making us use computer facilities.
This research was partly supported by the Grant-in-Aid for
Scientific Research from the Japanese Ministry of Education,
Science, and Culture under the contract number 06640406.

\vfill\eject
\vspace{1.0cm}
\noindent
{\Large\bf Appendix A. Running coupling constant}
\vspace{0.4cm}

We give explicit expressions of the running coupling constant.
It is given as
$$
\alpha_s^{LO}(Q^2) = {4\pi \over \beta_0 \ln (Q^2/\Lambda^2)} ~~~,
\eqno{(A.1)}
$$
in the leading order (LO), and it is
$$
\alpha_s^{NLO}(Q^2)={4\pi \over \beta_0 \ln(Q^2/\Lambda^2)}
\Biggl[1-{\beta_1 \ln \ln(Q^2/\Lambda^2) \over \beta_0^2\ln(Q^2/\Lambda^2)}
\Biggr]    ~~~,
\eqno{(A.2)}
$$
in the next-to-leading-order (NLO) \cite{RGR}.
The renormalization scheme is
$\overline{MS}$.
$\Lambda$ is the QCD scale parameter which depends on
the renormalization scheme .
$\beta_0$ and $\beta_1$ are the expansion coefficients
of the $\beta$-function and they are given by
$$
\beta_0={11\over3}C_G-{4 \over 3}T_R N_f  ~~~,~~~
\beta_1={34\over3}C^2_G-{10\over3}C_G N_f-2C_F N_f  ~~~,
\eqno{(A.3)}
$$
where $C_G$, $C_F$, and $T_R$ are constants associated with
the color $SU(3)$ group:
$$
C_G=N_c~~~,~~~ C_F={N_c^2-1 \over 2N_c}~~~,~~~ T_R={1 \over2} ~~~.
\eqno{(A.4)}
$$
$N_c$ is the number of color ($N_c$=3) and
$N_f$ is the number of flavor.

\vspace{1.0cm}
\begin{tabbing}
{\Large\bf Appendix B.} $~$ \= {\Large\bf Splitting function}       \\
$~~~$                       \> {\Large\bf and coefficient function}
\end{tabbing}
\vspace{0.4cm}

Spin-independent flavor-nonsinglet
splitting functions and
the coefficient functions
for calculating F$_1$, F$_2$, and F$_3$
are given in the following.
The LO nonsinglet splitting function is given by
$$
P_{NS}^{(0)}(x)=
C_F\biggl[{{1+x^2}\over {(1-x)_+}}+{3\over2}\delta(x-1)\biggr] ~~~,
\eqno{(B.1)}
$$
where the integral including the function $1/(1-x)_+$ is defined by
$$
\int dx{f(x) \over (1-x)_+}=\int dx {f(x)-f(1) \over 1-x}  ~~~.
\eqno{(B.2)}
$$
The NLO nonsinglet splitting function for a ``$q+\bar q$ type"
distribution [$\displaystyle{q_{_{NS}}=\sum_i \alpha_i (q_i+\bar q_i)}$]
is given by Herrod and Wada \cite{HW}
\begin{eqnarray*}
P_{NS+}^{(1)}(x)&=&C_F^2\biggl\{P_F(x)+P_A(x)+\delta(1-x)
\biggl[{3\over8}-{1\over2}\pi^2
+\zeta(3)-8\widetilde{S}(\infty)\biggr]\biggr\}  \\
&+&{1\over2}C_FC_A\bigg\{P_G(x)-P_A(x)+\delta(1-x)\biggl[{17\over12}+{11\over9}
\pi^2-\zeta(3)+8\widetilde{S}(\infty)\biggr]\biggr\}  \\
&+&C_FT_RN_f\biggl\{P_{N_F}(x)-\delta(1-x)\biggl({1\over6}+{2\over9}\pi^2\biggr)
\biggr\}
\end{eqnarray*}
\vspace{-1.5cm}

\hfill{(B.3)}

\vspace{+0.5cm}
\noindent
where $P_F(x)$, $P_G(x)$, $P_{N_F}(x)$, and $P_A(x)$ are given
by Curci, Furmanski, and Petronzio \cite{HW} as
\begin{eqnarray*}
P_F(x)\!\!&=&\!\!-2{{1+x^2} \over {1-x}}\ln x\ln(1-x)-
\biggl[{3\over {1-x}}+2x\biggr]\ln x-{1\over2}(1+x)\ln^2x  -\!\!5(1-x)   ~,\\
P_G(x)\!\!&=&\!\!{1+x^2\over (1-x)_+}\bigg[\ln^2x+{11 \over3}\ln x+{67\over9}
-{1\over3}\pi^2\biggr]+2(1+x)\ln x+{40\over 3}(1-x)                     ~~~, \\
P_{N_F}(x)\!\!&=&\!\!{2\over 3}\Biggl[{1+x^2\over (1-x)_+}\biggl(
-\ln x-{5\over 3}\biggr)-2(1-x)\Biggr]                                  ~~~, \\
P_A(x)\!\!&=&\!\!2{1+x^2\over 1+x}\int_{x/(1+x)}^{1/(1+x)}{dz \over z}
\ln {{1-z} \over z}
+2(1+x)\ln x+4(1-x)                                                     ~~~.
\end{eqnarray*}
\vspace{-1.5cm}

\hfill{(B.4)}

\vspace{+0.5cm}
\noindent
The $\zeta$ function is defined by
$\displaystyle{\zeta (k)=\sum_{n=1}^\infty {1 \over n^k}}$
and the numerical value ($\zeta(3)$=1.2020569...)
is taken from Ref. \cite{AS}. $\widetilde S(\infty)$ is given
by the $\zeta$ function as
$ \widetilde S (\infty) =-{5 \over 8}\zeta(3) $
in the Gonzalez-Arroyo paper \cite{MELLIN}.
In the case of a ``$q-\bar q$ type", we simply change the sign
of $P_A$ ($P_A\rightarrow -P_A$) \cite{HW}.
\begin{eqnarray*}
P_{NS-}^{(1)}(x)&=&C_F^2\biggl\{P_F(x)-P_A(x)+\delta(1-x)
\biggl[{3\over8}-{1\over2}\pi^2
+\zeta(3)-8\widetilde{S}(\infty)\biggr]\biggr\}  \\
&+&{1\over2}C_FC_A\bigg\{P_G(x)+P_A(x)+\delta(1-x)\biggl[{17\over12}+{11\over9}
\pi^2-\zeta(3)+8\widetilde{S}(\infty)\biggr]\biggr\}  \\
&+&C_FT_RN_f\biggl\{P_{N_F}(x)-\delta(1-x)\biggl({1\over6}+{2\over9}\pi^2\biggr)
\biggr\}     ~~~.
\end{eqnarray*}
\vspace{-1.5cm}

\hfill{(B.5)}

\vspace{+0.5cm}

The coefficient functions are needed in
calculating structure functions with the NLO corrections.
The coefficient functions $B^{NS} (x)$ for the structure
functions, $F_1$, $F_2/x$, and $F_3$, are
\begin{eqnarray*}
B_1^{NS}(x)&=&{C_F \over 2}[F_q(x)-4x]  ~~~, \\
B_2^{NS}(x)&=&C_F F_q(x)                ~~~, \\
B_3^{NS}(x)&=&C_F [F_q(x)-2-2x]         ~~~,
\end{eqnarray*}
\vspace{-1.5cm}

\hfill{(B.6)}

\vspace{+0.5cm}
\noindent
where the function $F_q(x)$ is given by
$$
F_q(x)=-{3 \over 2}{1+x^2 \over (1-x)_+}+{1\over 2}(9+5x)-2{1+x^2 \over 1-x}
\ln x +2(1+x^2)\biggl[ {\ln (1-x) \over 1-x} \biggr]_+
                  -\delta (1-x)(9+{2 \over 3}\pi^2) ~~.
\eqno{(B.7)}
$$

\vspace{1.0cm}
\begin{tabbing}
{\Large\bf Appendix C.} $~$ \= {\Large\bf Laguerre polynomials} \\
$~~~$                       \> {\Large\bf and the LO evolution operator}
\end{tabbing}
\vspace{0.4cm}

We use the Laguerre polynomials defined by \cite{AS}
$$
L_n(x^\prime)=\sum_{m=0}^n (-1)^m \left( \begin{array}{c}
                                          n \\ m
                                         \end{array}     \right)
                  {{x^{\prime~m}} \over {m!}}
{}~~.
\eqno{(C.1)}
$$
It should be noted that a slightly different
definition, the above equation multiplied by $n!$,
is sometimes used. A function of the variable $x$ is first expressed
as the function of $x^\prime=-\ln x$, then it is expanded by
the Laguerre polynomials.
$$
f (x=e^{-x^\prime})=\sum_n f_n L_n(x^\prime) ~~~,
\eqno{(C.2)}
$$
where $f_n$ is the expansion coefficient given by the integral
$$
f_n \equiv \int_0^\infty dx^\prime e^{-x^\prime} L_n(x^\prime)
f (e^{-x^\prime})  ~~~.
\eqno{(C.3)}
$$
This integral is calculated numerically by a Gauss quadrature.
If the initial distribution can be expressed
in a simple analytical form $f(x)=ax^b(1-x)^c$, the coefficients are
obtained by \cite{RAM}
$$
f_n = a \sum_{k=0}^\infty g_k(c)  {{(k+b)^n} \over {(k+b+1)^{n+1}}}
{}~~,
\eqno{(C.4)}
$$
with the recurrence relations $g_0(c)=1$, ...,
$g_{k+1}(c)=-g_k(c)(c-k)/(k+1)$.

The LO evolution operator is expanded by the Laguerre polynomials,
and the method of calculating the expansion coefficients are discussed
in Refs. \cite{FP,RAM,KL}. We only show the results in the following.
The coefficient is given by
$$
E_m^{(0)}(t)=e^{P_0t}\sum_{k=0}^N{t^k\over k!}B_m^k ~~~,
\eqno{(C.5)}
$$
where $B_m^k$ is calculated by the recurrence relation
$$
B_m^{k+1}= \sum_{i=k}^{m-1} (P_{m-i}-P_{m-i-1}) B_i^k   ~~~,
\eqno{(C.6)}
$$
with
$$
B_i^0=1,\ B_i^1=\!\sum_{j=1}^i (P_j-P_{j-1}),\ B_0^k=B_1^k= \cdots
=B_{k-1}^k=0    ~~~.
\eqno{(C.7)}
$$
Namely, we first calculate the expansion coefficients of
the splitting function ($P_n$), then $B_m^k$ is calculated
by the recurrence relation in Eqs. (C.6) and (C.7).

\vfill\eject

\vfill\eject
\noindent
{\Large\bf{Figure Captions}} \\

\vspace{-0.38cm}
\begin{description}
   \item[Fig. 1]
Valence-quark distributions calculated by our program LAG2NS.
The valence-quark distribution at $Q^2$=4 GeV$^2$ is
the HMRS-B distribution, which is
evolved to the ones at $Q^2$=200 GeV$^2$.
Three results are shown; the dotted, dashed, and solid curves
are obtained by taking ten, twenty, and thirty Laguerre polynomials
respectively. See text for the details.
   \item[Fig. 2]
Our evolution results are compared with CDHSW experimental
data. We choose the CCFR distribution at $Q^2$=3 GeV$^2$
in both LO and NLO cases, then we calculated $Q^2$ variations.
$F_3$ structure functions at $x$=0.045, 0.225, and 0.55
are shown.
The solid (dashed) curves are the results of the NLO (LO) $Q^2$ evolution.
\end{description}

\vfill\eject
\noindent
{\Large\bf{TEST RUN OUTPUT}} \\
\vspace{0.4cm}

\noindent
****** ENTER ****** F0,IORDER,ITYPE,IMORP,Q2,Q02,DLAM,   \par\noindent
{}~~~~~~~~~~~~~~~~~~~~~~~~~~~~~~~NMAX,IQN,ILOG,XMIN,XMAX,NXSTEP   \par\noindent
F0= 4.0~IORDER= 2~ITYPE= 4~IMORP= 1~Q02= 4.0~Q2=200.000~DLAM= 0.190
\par\noindent
NMAX=30~IQN=1~ILOG=2~XMIN= 0.0100~XMAX= 1.00~NXSTEP= 30   \par\noindent
\begin{tabular}{cc}
     $~~~~~$X$~~~~~$       &    $~~~~~$XFQNS1$~~~~~$     \\
       0.010000  &  0.2627824   \\
       0.011659  &  0.2799758   \\
       0.013594  &  0.2989168   \\
       0.015849  &  0.3196428   \\
       0.018478  &  0.3421817   \\
       0.021544  &  0.3665633   \\
       0.025119  &  0.3928285   \\
       0.029286  &  0.4210298   \\
       0.034145  &  0.4512194   \\
       0.039811  &  0.4834202   \\
       0.046416  &  0.5175738   \\
       0.054117  &  0.5534666   \\
       0.063096  &  0.5906320   \\
       0.073564  &  0.6282371   \\
       0.085770  &  0.6649641   \\
       0.100000  &  0.6989059   \\
       0.116591  &  0.7275024   \\
       0.135936  &  0.7475522   \\
       0.158489  &  0.7553388   \\
       0.184785  &  0.7469143   \\
       0.215443  &  0.7185751   \\
       0.251189  &  0.6675502   \\
       0.292864  &  0.5928846   \\
       0.341455  &  0.4964421   \\
       0.398107  &  0.3838530   \\
       0.464159  &  0.2650883   \\
       0.541170  &  0.1541317   \\
       0.630957  &  0.0669216   \\
       0.735642  &  0.0163349   \\
       0.857696  &  0.0024327   \\
       1.000000  & --0.0045267   \\
\end{tabular}


\begin{thebibliography}{}
\bibitem{MISTAKE} The $\beta_0/2$ factor in Eq. (3.26) of Ref. \cite{FP}
                  should be corrected to $2/\beta_0$.
\end{thebibliography}
\end{document}